\documentclass[preprint,showpacs,prl,preprintnumbers,amsmath,amssymb]{revtex4}
\usepackage{graphicx}
\usepackage{bm}
\usepackage{verbatim}%\begin{comment} mult lines of comment \end{comment}

\begin{document}
%\preprint{LA-UR }
\title{Anisotropic small-scale constraints on energy in rotating stratified turbulence}
\author{Susan Kurien$^\dag$, Beth Wingate$^\ddag$ and Mark A. Taylor$^*$}
\affiliation{\dag Theoretical Division,~Los Alamos National Laboratory, Los
  Alamos, NM 87545, USA\\
\ddag Computer and Computational Sciences,~Los Alamos National Laboratory, Los Alamos, NM 87545, USA\\
* Exploratory Simulations Technologies,
  Sandia National Laboratories, Albuquerque, NM 87185, USA} 
\date{\today}

\begin{abstract}
  Rapidly rotating, stably stratified three-dimensional inviscid flows
  conserve both energy and potential enstrophy.  We show that in such
  flows, the forward cascade of potential enstrophy imposes
  anisotropic constraints on the wavenumber distribution of kinetic
  and potential energy. The horizontal kinetic energy is suppressed in
  the large, nearly horizontal wave modes, and should decay with the
  horizontal wavenumber as $k_h^{-3}$. The potential energy is
  suppressed in the large, nearly vertical wave modes, and should
  decay with the vertical wavenumber as $k_z^{-3}$. These results
  augment the only other exact prediction for the scaling of energy
  spectra due to constraints by potential enstrophy obtained by
  Charney (J.  Atmos. Sci. 28, 1087 (1971)), who showed that in the
  quasi-geostrophic approximation for rotating stratified flows, the
  energy spectra must scale isotropically with total wavenumber as
  $k^{-3}$. We test our predicted scaling estimates using resolved
  numerical simulations of the Boussinesq equations in the relevant
  parameter regimes, and find reasonable agreement.
\end{abstract}


\maketitle
%\section{Introduction}

Classical quasi-geostrophic (QG) flow is a useful approximation for
rapidly rotating, strongly stratified flows \cite{Pedlosky,Salmon}. In this
approximation, the zeroth-order expansion of the velocity in the
rotation and stratification parameters is geostrophic, meaning that
the Coriolis force is balanced by the pressure gradient force.
Furthermore the linear plane-waves called
inertia-gravity waves are eliminated to the lowest order. This leads
to a simplification of the dynamics which is described entirely by the
evolution of potential vorticity $q_{qg}$ \cite{Charney71}:
\begin{eqnarray}
  \frac{\partial q_{qg}}{\partial t} &+& {\bm{u}_0}_h \cdot \nabla q_{qg} = 0, \label{qgpvdyn}\\
  \mbox{where}\quad q_{qg} &=& f \frac{\partial \theta}{\partial z} - N{\omega}_3,
  \label{charneypv}
\end{eqnarray}
where ${\bm{u}_0}_h$ is the leading order (horizontal) geostrophically
balanced velocity, $\theta$ is the density fluctuation scaled to have
the same dimensions as velocity, $\omega_3 = (\nabla_h \times
{\bm{u}}_h)\cdot \hat{\bm{z}}$ is the $z$-component of the vorticity,
$f$ and $N$ are the Coriolis and Brunt-V\"ais\"al\"a (buoyancy)
frequencies respectively of a system which is rotating stratified
in the $z$-direction. In QG flow, the vertical velocity $\bm{w}$
appears as a correction to leading order geostrophic balance.

In 1971, Charney \cite{Charney71} showed that the global conservation
of total energy $E_T = \frac{1}{2}\int(|\bm{u}|^2 + \theta^2) d\bm{x}$, and
potential enstrophy $Q_{qg} = \frac{1}{2}\int |q_{qg}|^2 d\bm{x}$ by inviscid
three-dimensional (3d) QG flow is analogous to conservation of energy
and enstrophy in non-rotating two-dimensional turbulence (see for
example \cite{KKTung01} for further discussion on the assumptions and details
of Charney's work).
Following the classical theory of 2d turbulence
\cite{Fjortoft53,Kr67}, Charney predicted an inverse (upscale) cascade
of energy with corresponding large-scale energy spectrum $E_T(k)
\propto \varepsilon^{2/3} k^{-5/3}$, and a forward (downscale) cascade
of potential enstrophy with corresponding small-scale energy spectrum
$E_T(k) \propto \varepsilon_Q^{2/3} k^{-3}$, where $k$ is the wavenumber and
$\varepsilon$ and $\varepsilon_Q$ are the transfer rates of energy and
potential enstrophy, respectively. 

In theory, as the rotation and stratification of a 3d fluid become
infinitely strong, the inertia-gravity waves are eliminated to lowest
order, giving leading order QG flow satisfying
(\ref{qgpvdyn},\ref{charneypv}).  In practice, for very large but
finite rotation and stratification, the inertia-gravity waves
strongly influence the small-scale dynamics leading to $E_T(k) \sim
k^{-\gamma}$ where $1 < \gamma < 2$ \cite{B95,BMNZ97} in the high
wavenumbers. The underlying leading order QG scaling of $k^{-3}$
predicted by Charney can then only be extracted by separating the QG (or
geostrophic) modes from the wave (or ageostrophic) modes by either
suitably projecting the full solutions onto the QG modes
\cite{B95,SW02} or by filtering out the ageostrophic inertia-gravity
waves \cite{Salmon}.

In the present work we consider rotating stratified
turbulence retaining both the leading order QG as well as all
sub-leading contributions from inertia-gravity waves and other
nonlinear waves. We show that in this parameter regime, as in
classical QG, potential enstrophy plays a significant role in
constraining the energy. For a wavevector $\bm{k} = k_x \hat{\bm{x}}+
k_y \hat{\bm{y}}+k_z \hat{\bm{z}}$ the horizontal component is $k_h =
(k_x^2 + k_y^2)^{1/2}$ and the vertical component is $k_z$. We show
that the potential enstrophy dominates over potential energy in the
large, nearly vertical modes ($k_z/k_h \gg 1$), resulting in a
potential energy spectrum scaling of $k_z^{-3}$ for large $k_z$. And
potential enstrophy also dominates over horizontal kinetic energy in
the large, nearly horizontal modes ($k_h/k_z \gg 1$), resulting in a
horizontal kinetic energy spectrum scaling of $k_h^{-3}$ for large
$k_h$. These are the first scaling estimates for the spectra of
rapidly rotating and stably stratified flows away from pure QG,
obtained solely using the relationship between potential enstrophy and
energy as a function of scale. Such scaling laws, apart from being benchmarks, 
are potentially
useful in parameterizing the turbulent small scales in large simulations of
 rotating
stratified flows thus reducing
the computational expense of explicitly resolving the small scales.

%\section{Near quasi-geostrophic dynamics and spectra }
We begin with the Boussinesq
equations for rotating, stably stratified and incompressible flow
given by \cite{EM98}:
\begin{eqnarray}
\frac{D}{Dt}\bm{u} + f\hat{\bm z}\times \bm{u} +
\nabla p + N\theta \hat{\bm{z}} &=& \nu\nabla^2 \bm{u} + {\cal F}\nonumber\\
\frac{D}{Dt}\theta - Nw &=& \kappa \nabla^2\theta \label{bous}\\
\nabla\cdot\bm{u} &=& 0,\nonumber
\end{eqnarray}
where $\displaystyle\frac{D}{Dt} = \frac{\partial}{\partial t} +
\bm{u}\cdot\nabla$, $\bm{u}$ is the velocity, $w$ is its vertical
component, $p$ is the effective pressure and $\cal F$ is an external
input or force.  The total density is ${\cal \rho}_T(\bm{x}) = \rho_0
- bz + \rho(\bm{x})$, where $\rho_0$ is the constant background, $b$
is also constant and larger than zero for stable stratification in the
vertical $z$-coordinate, $\rho$ is the density fluctuation such that
$| \rho | \ll | bz | \ll \rho_0$ and $\displaystyle \theta =\rho
({g}/{b \rho_0})^{1/2}$ has the dimensions of velocity. The Coriolis
parameter $f = 2\Omega$ where $\Omega$ is the constant rotation rate
about the $z$-axis, the Brunt-V\"ais\"al\"a frequency $\displaystyle N
= ({gb}/{\rho_0})^{1/2}$, $\nu=\mu/\rho_0$ is the kinematic viscosity
and $\kappa$ is the mass diffusivity coefficient. We assume periodic
or infinite boundary conditions.  The relevant non-dimensional
parameters for this system are the Rossby number $Ro = f_{nl}/{f}$ and
the Froude number $Fr = f_{nl}/{N}$, where $f_{nl} = (\epsilon_f
k_f^2)^{1/3}$ is the non-linear frequency given input rate of energy
$\epsilon_f$ \cite{SW02}. Thus $Ro$ and $Fr$ are the ratios of
rotation and stratification timescales respectively to the nonlinear
timescale.

The Boussinesq equations conserve the following quantities for
 ${\cal F} = \nu=\kappa =0$, 
\begin{eqnarray}
  \mbox{total energy}~E_T = E + P,~~
\frac{ D}{D  t}\int E_T~d\bm{x} &=& 0,\nonumber\\
\mbox{potential vorticity}~q = \Big(\bm{\omega}_a \cdot \nabla
\rho_T\Big),~~
 \frac{D q}{D t}&=& 0, \nonumber \\
  \mbox{potential enstrophy}~ Q =\frac{1}{2}q^2,~~
\frac{D Q}{D t}= \frac{D}{D t}\int Q~d\bm{x}&=& 0.\nonumber
\end{eqnarray}
$E=\frac{1}{2}|\bm{u}|^2$ is the kinetic energy,
$P=\frac{1}{2}\theta^2$ is the potential energy of the density
fluctuations. The
absolute vorticity $\bm{\omega}_a = \bm{\omega} + f\hat{\bm{z}}$ and
the relative (or local) vorticity $\bm{\omega} = \nabla \times
\bm{u}$. Potential vorticity may be written in terms of $\theta$ as
\begin{equation}
 q = fN + \bm{\omega}\cdot\nabla\theta + f  \frac{\partial \theta}{\partial z} - N\omega_3.
\label{pv_theta}
\end{equation}
The constant part $fN$ does not participate in the dynamics and we will therefore neglect it from now on. 
The linear part of (\ref{pv_theta}) is precisely $q_{qg}$
of (\ref{charneypv}). In what follows we will assume that $\nu
\rightarrow 0$ and $\kappa \rightarrow 0$ such that Prandtl number $Pr
= \nu/\kappa = 1$, and the force $\cal{F}$ is confined to the lowest
modes. Thus we assume a conventional `inertial-range' of turbulent
scales wherein the transfer of conserved quantities dominates over
both their dissipation and forcing.

\begin{comment}
In the limit $Ro \rightarrow 0$, $Fr \rightarrow 0$, the linear waves
known as inertia-gravity waves are eliminated in the leading order
solution, yielding classical QG as described by (\ref{charneypv}).
Equivalently, one could project out the linear solutions of
Eq.~(\ref{bous}) which carry the linear potential vorticity
\cite{B95,SW02}. Quasi-geostrophic modes are thus a subset of the full
solution to the Boussinesq equations in the limit $Ro \rightarrow 0$,
$Fr \rightarrow 0$, a limit which is difficult to achieve in practice.
\end{comment}

As $Ro \rightarrow 0 $ and $Fr \rightarrow 0$, the potential vorticity
$q$ approaches $q_{qg}$ \cite{B95,KSW06}. This is easily observed by
considering the non-dimensional form of (\ref{pv_theta}) namely $q =
\bm{\omega}\cdot\nabla\theta + Ro^{-1} \frac{\partial \theta}{\partial
  z} - Fr^{-1}\omega_3$, and letting $Ro$ and $Fr$ tend to zero
together. In fourier representation:
\begin{eqnarray}
  \tilde{q}(\bm{k})  \simeq f k_z \tilde{\theta} + i N \bm{k}_h \times
  \tilde{\bm{u}}_h = f k_z \tilde{\theta} + i N k_h \tilde{u}_h
  \label{qg_pvf}
\end{eqnarray}
where $\tilde{\cdot}$ denotes fourier coefficients, the total
wavevector $\bm{k}$ = $\bm{k}_h + k_z \hat{\bm{z}}$, the horizontal
wavevector component has length $k_h = ({k_x^2 + k_y^2})^{1/2}$, the
vertical wavenumber is $k_z$ and $\bm{u}_h$ is the horizontal velocity
vector with magnitude $u_h = (u_x^2 + u_y^2)^{1/2}$. We assume that
the vertical velocity $w = u_z \sim 0$ in the lowest order (classical
QG) thus obtaining the last equality of Eq.~(\ref{qg_pvf}). 

We take both $N$ and $f$ to be very large, and $N/f = 1$ so that $Ro =
Fr$. This approaches the special case $Ro \rightarrow 0$ and $Fr
\rightarrow 0$ while $Ro = Fr$ which was shown rigorously to be
leading order QG in \cite{EM98}. For $k > k_f$, scales
smaller than the forcing scale, we consider two cases:

\noindent
{\it  1) Case $\displaystyle \frac{k_z}{k_h} \gg 1$ \label{case 1}}.
These are the more vertical wavenumber modes corresponding loosely to 
flat `pancake' scales in physical space.
Eq.~(\ref{qg_pvf}) reduces to $\tilde{q}  \simeq f k_z \tilde\theta$, 
yielding the following relation between potential enstrophy and potential energy distribution in spectral space,
\begin{equation}
  Q(k_h, k_z) = 
  \frac{1}{2}|\tilde{q}|^2 = f^2  k_z^2 P(k_h,
  k_z),\label{qg1_Q}\nonumber
\end{equation}  
which upon integration over some high vertical wavenumber interval leads to the following constraint:
\begin{equation}
\int_{\kappa_z}^\infty Q(k_h,  k_z) dk_z \gg  
  f^2 \kappa_z^2 \int_{\kappa_z}^\infty  P(k_h,k_z)~dk_z,
\label{qg2_Q}\nonumber
\end{equation}
where the potential energy spectrum
$P(k_h,k_z)=\frac{1}{2}|\tilde\theta|^2$.
Thus, for sufficiently high wavenumbers $\kappa_z \rightarrow \infty$,
the potential enstrophy $Q$ forms the dominant forward cascade and, in
order to remain finite, suppresses the potential energy $P$ in this
regime. The dimensional argument
following \cite{Fjortoft53,Kr67} assumes that in this wavenumber limit,
the potential energy spectrum must depend on the potential enstrophy flux rate
$\varepsilon_Q$ and the vertical wavenumber $k_z$, so that:
\begin{equation} P(k_h, k_z) \sim  
\varepsilon_Q ^{2/5} k_z^{-3}.
\label{Pscaling}
\end{equation} 
\begin{comment}
For completeness, we can also propose an inverse cascade of
potential energy toward small $k_z$ for small $k_h$, depending only on the
energy dissipation rate. Then, we must have that up to a constant
\begin{equation}
  P(k_h, k_z) \sim \varepsilon^{2/3}k_z^{-5/3}
\end{equation}
\end{comment}

\noindent
{\it 2) Case $\displaystyle \frac{k_h}{k_z} \gg 1 $ \label{case2}}.
These are the wide flat wavenumber modes corresponding to 
the tall columnar scales in physical space. 
In this limit Eq.~(\ref{qg_pvf}) reduces to
$\tilde{q}  = i N k_h \tilde{u}_h$.
Following the same arguments as for potential energy above, 
we obtain that the potential enstrophy dominates the forward cascade
in the regime $k_h/k_z \gg 1$, resulting in suppression of horizontal kinetic energy resulting in the following scaling estimate:
\begin{equation}E_h(k_h,k_z) \sim \varepsilon_Q ^{2/5} k_h^{-3}.
\label{Escaling}
\end{equation} 
where the horizontal kinetic energy $E_h(k_h,k_z)= \frac{1}{2}|\tilde{u}_h|^2$.

Our new scaling predictions (\ref{Pscaling}) and (\ref{Escaling}) 
describe the spectral statistics of the full
flow for small but finite $Ro$ and $Fr$, not just the leading order QG
part of the dynamics. The main purpose of
this paper is to show that the potential enstrophy imposes predictable
constraints on the energy in rapidly rotating stratified flow {\it
  even when the flow is not strictly QG.}  In the next sections we
seeking to numerically verify our predictions for the energy spectra in the two limiting regimes in wavenumber.

%\section{Numerical simulations}
We simulate the Boussinesq equations (\ref{bous}) taking $N = f$ very
large. We use a pseudo-spectral code in a periodic cube of side $L =
1$, generating wavenumbers which are integer multiples of $2\pi$. A
fourth-order Runge-Kutta time integration is used and the
inertia-gravity wave frequencies are resolved in our explicit scheme.
Since we are interested in the small scales (high wavenumbers), we use
a low-wavenumber stochastic forcing centered at $k_f = 4$. In order to
extend the inertial range in the high wavenumbers, the viscous
dissipation is modeled using a hyperviscous term $(-1)^{p
  +1}\nu(\nabla^2)^p \bm{u}$, where $p = 8$ in place of the normal
laplacian viscosity term $\nu \nabla^2\bm{u}$.  The hyperviscosity
coefficient $\nu$ is dynamically chosen based on the energy in the
highest mode for both momentum and mass diffusion following
\cite{SW02}, $\nu(t) = 2.5 \Big (\frac{E(k_m,t)}{k_m}\Big)^{1/2}
k_m^{2-2p}$ where $k_m$ is the highest available wavenumber and
$E(k_m,t)$ is the kinetic energy in that wavenumber. An analogous
scheme is used for the diffusion term in Eq.~(\ref{bous}) for the
evolution of $\theta$.  The parameters of a few of our runs are given in
Table \ref{table}. We report the results from data $\#4$ for which
$Ro$ and $Fr$ are the smallest and the resolution is the best.
\begin{table}
\centering
\begin{tabular}{ccccccc}
  \hline
\#&  $n$ & $k_f$ & $N/f$&$Ro$ & $Fr$ & $\epsilon_f$ \\
  \hline  
1&  256 & 4 & 1 &0.029 & 0.029 & 0.62 \\
2&  256 & 4 &1  & 0.014 & 0.014 & 0.61 \\
3&  256 & 4 & 1& 0.0072 & 0.0072  & 0.60\\
4&  512 & 4 & 1& 0.0072 & 0.0072 & 0.60\\
\hline
\end{tabular}
\caption{Parameters of Boussinesq calculations: $n$ -- number of grid points to a
  side; $k_f$ -- forcing wavenumber; $N$ -- Brunt-V\"ais\"al\"a frequency,
  $Ro$ -- Rossby number; $Fr$ -- Froude number; $\epsilon_f$ -- rate
  of input of kinetic energy. }\label{table}
\end{table}

\begin{figure}[ht]
\centering
\includegraphics[width=2.75in]{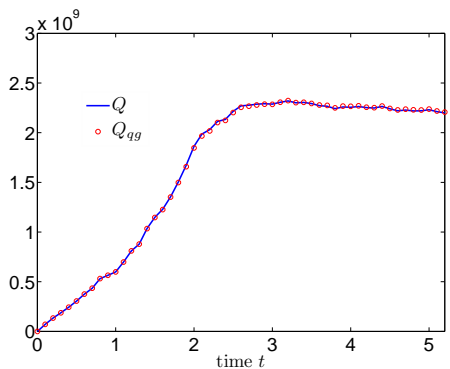}
\caption{Total potential enstrophy $Q$ and its linear part $Q_{qg}$
  for data \#4.  The two are indistinguishable, indicating purely
  quadratic potential enstrophy.  Time $t$ is in dimensional units. }
\label{qg_potens}
\end{figure}
Figure \ref{qg_potens} shows the evolution of the total potential
enstrophy $Q$ from (\ref{pv_theta}), and its linear, quasi-geostrophic
piece $Q_{qg}$.  The former is indistinguishable from the latter
indicating that the nonlinear part of the potential vorticity $\bm{\omega}\cdot
\nabla\theta$ is negligible and thus the potential enstrophy is
quadratic. The system is QG in the leading order, or near-QG in the
sense described above. The mean potential enstrophy has
reached a nearly steady value in the time range $2.5 < t < 5.2$,
which corresponds to between 5 and 11 non-linear time cycles, or 5000 to
10400 rotation (stratification) cycles.

\begin{figure}[ht]
\centering
\includegraphics[width=2.75in]{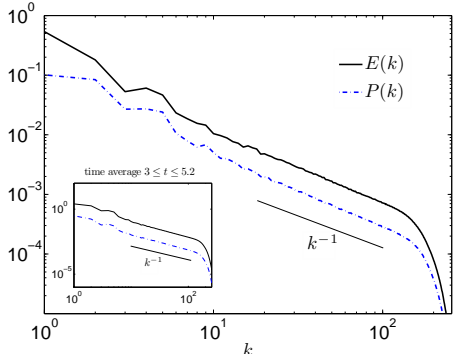}
\caption{Log-log plot of spherical shell averaged potential and
  kinetic energy spectra for data $\#4$ at time $t = 5.2$. The high
  wavenumber scaling is $k^{-1}$ indicating that in this
  representation the energy is dominated by waves. Inset: Same spectra averaged
  over time $3\leq t \leq 5.2$.}
\label{EkPk}
\end{figure}
Figure \ref{EkPk} shows the shell-averaged kinetic and potential
energy spectra for our simulation, computed as follows:
\begin{eqnarray}
  E(k) = \frac{1}{2}\sum_{k'} |\tilde{\bm{u}}(\bm{k}')|^2, ~~
  P(k) = \frac{1}{2}\sum_{k'} |\tilde\theta(\bm{k}')|^2 \nonumber
\end{eqnarray}
where $k-0.5 \leq k' < k+0.5$ thus including all wavenumbers in the
spherical shell of average radius $k$. The scaling of both $E(k)$ and
$P(k)$ is $k^{-1}$ for $k \gg k_f$ which indicates that by this
measure the high wavenumbers are still dominated by waves
\cite{B95,SukSmi07}. The inset shows the average of the spectra over
the time period $3 \leq t \leq 5.2$ over which the potential enstrophy
is constant as seen in Fig. \ref{qg_potens}. The time-averaged spectra
also show a scaling very close to $k^{-1}$ indicating that the small
scales ($k > k_f$) have achieved close to a statistically steady
state.

\begin{figure}[ht]
\centering
\includegraphics[width = 2.75in]{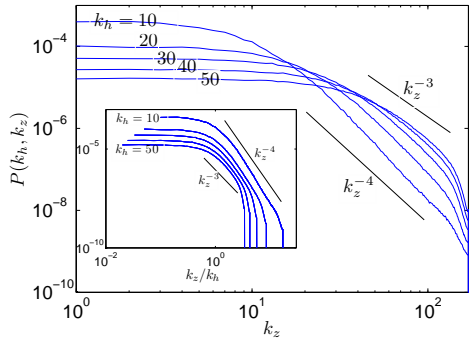}
\caption{Log-log plot of potential energy density $P(k_h,k_z)$ vs. $k_z$ for
  data $\#4$ averaged over time $3 \leq t \leq 5.2$. Each curve is the
  spectrum for a different fixed value of $k_h$. For $10 \leq k_h
  \leq 50$ and $k_z \gg k_f$, the scaling ranges between $k_z^{-4}$
  and $k_z^{-3}$. Inset: Same spectra
  versus $k_z/k_h$ shows that the `turnover' to the inertial range
  scaling for all the curves emerges only when $k_z/k_h \geq 1$.
}
\label{qg_Pkhkz}
\end{figure}
The potential energy and horizontal kinetic energy spectra as
functions of $k_h$ and $k_z$ were computed as double sums according to:
\begin{equation}
P(k_h,k_z) = \frac{1}{2}\sum_{k_h',k_z'}|\tilde\theta(\bm{k}')|^2,~~E_h(k_h,k_z) = \frac{1}{2}\sum_{k_h',k_z'}|\tilde{\bm{u}}_h(\bm{k}')|^2\nonumber 
\end{equation} where $k_z[k_h] - 0.5 \leq k_z'[k_h'] < k_z[k_h] + 0.5$.
Figure \ref{qg_Pkhkz} shows $P(k_h,k_z)$ as a
function of $k_z$ for various values of $k_h$. For $10\leq k_h \leq 50$ and
$k_f \leq k_z \leq 100 $, the scaling for $P(k_h, k_z)$ ranges between $k_z^{-4}$ and $k_z^{-3}$ indicating stronger suppression of potential energy 
than the dimensional prediction of
Eq.~(\ref{Pscaling}). As $k_h$ increases $P(k_h,k_z)$
also persists more strongly into the high $k_z$. Conversely, for a
fixed small $k_z \leq k_f$, the smaller $k_h$ spectra have more
energy, indicative of a growth of potential energy as $k_z \ll k_f$
for small $k_h$. The inset of Fig.~\ref{qg_Pkhkz} shows the same
spectra versus $k_z/k_h$ which shows clearly that the inertial
range scaling for each $k_h$ emerges only when $k_z/k_h \geq 1$, the
predicted range for Eq.~(\ref{Pscaling}). Overall the constraints on
potential energy due to potential enstrophy are thus highly
anisotropic in scale and consistent with our prediction. 

\begin{figure}[ht]
\centering
\includegraphics[width=2.75in]{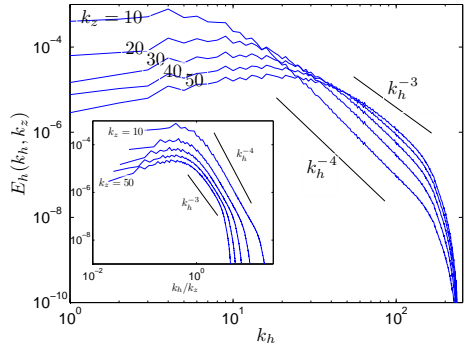}
\caption{Log-log plot of horizontal kinetic energy density $E_h(k_h,k_z)$ vs.
  $k_h$ for data $\#4$ averaged over time $3 \leq t \leq 5.2$.  Each
  curve is a different value of $k_z$. For $10 \leq k_z \leq 50$ there
  the scaling ranges between $k^{-4}$ and $k^{-3}$. Inset: Same
  spectra vs. $k_h/k_z$ showing that the inertial range emerges as
  $k_h/k_z \geq 1$ as predicted. }
\label{qg_Ehkhkz}
\end{figure}
Figure \ref{qg_Ehkhkz} shows $E_h{k_h,k_z}$ as a function of $k_h$ for various
values of $k_z$. For $10 \leq k_z
\leq 50$ and $10 < k_h < 100$, the horizontal energy spectrum
$E_h(k_h,k_z)$ scales between $k_h^{-4}$ and $k_h^{-3}$ consistent
with the suppression of horizontal kinetic energy by potential
enstrophy in these modes. As $k_h$ grows, the horizontal kinetic
energy persists more strongly into the high $k_h$. Conversely, for a
fixed small $k_h$, the smaller $k_z$ have more energy, indicating a
growth of energy upscale in $k_z$. The inset of Fig.~\ref{qg_Ehkhkz}
shows again that the inertial range scaling predicted arises only in the 
anisotropic regime $k_h/k_z \geq 1$ consistent with prediction.

%\section{Conclusions }
In conclusion, we have deduced separate scaling laws for horizontal
kinetic and potential energy spectra, both of which are constrained in
the small scales by potential enstrophy in rapidly rotating stably
stratified flows. Potential enstrophy suppresses the potential energy
in the large, nearly vertical modes, and also suppresses horizontal
kinetic energy in the large, nearly horizontal modes; the resulting
energy densities in these modes scales as $k_z^{-3}$ and $k_h^{-3}$
respectively. Our test simulations data show even steeper scaling of
the spectra than predicted (greater suppression due to potential
enstrophy). The numerical calculations used to verify our predictions
are, at 512 grid-points to a side, the highest resolution unit
aspect-ratio simulations of the Boussinesq equations performed to
date. Higher resolution may well show closer agreement with our
theoretical prediction; we have already observed a tendency toward our
predicted $-3$ exponent when going from $256^3$ to $512^3$ in
grid-resolution. Most importantly, the scalings predicted and observed
are very different from the isotropic $k^{-\gamma}$ ($1< \gamma < 2 $)
scaling expected, for example, in the wave-dominated shell-averaged
spectra for near-QG flows (see Fig.~\ref{qg_potens}).  Our only
assumption is that rotation and stratification are strong enough that
the potential vorticity becomes linear, and hence the potential
enstrophy quadratic. We do not invoke additional asymptotics nor do we
need to limit ourselves only leading order modes. In future work we
will extend our analysis to the case of $N/f \neq 1$, that is, the
strength of rotation and stratification are large but unequal. The
possibilities for generalized quasi-geostrophic flow
\cite{JulienKMW06} in which aspect ratio is an additional parameter,
are also promising areas for further research.

We thank Leslie Smith, Jai Sukhatme and Greg Eyink for valuable
discussions during the course of this work. The work was funded by DOE
Office of Science Advanced Scientific Computing Research (ASCR)
Program in Applied Mathematics Research, the National Nuclear Security
Administration of the U.S.  Department of Energy at Los Alamos
National Laboratory under Contract No. DE-AC52-06NA25396, the
Laboratory Directed Research and Development program, and
NSF-DMS-0529596.

\bibliography{../../Bibs/jfm-pv}
\end{document}